\documentclass{ws-mpla_arXiv11122669} 

\usepackage{graphicx}                 
\DeclareRobustCommand\openone{\leavevmode\hbox{\small1\normalsize\kern-.33em1}}

\begin{document}

\title{NEUTRINO MASS AND THE STANDARD MODEL}

\author{F.R. Klinkhamer}
\address{Institute for Theoretical Physics, 
Karlsruhe Institute of Technology (KIT),\\
76128 Karlsruhe, Germany\\
frans.klinkhamer@kit.edu}

\maketitle
\begin{abstract}
It is pointed out (not for the first time)
that the minimal Standard Model, without additional gauge-singlet
right-handed neutrinos or isotriplet Higgs fields,
allows for non\-vanishing neutrino masses and mixing.
The required interaction term is nonrenormalizable and
violates $B-L$ conservation.
The ultimate explanation of this interaction term
may or may not rely on grand unification.%
\vspace*{1.25\baselineskip}\newline
Journal: \emph{Mod. Phys. Lett. A} 28, 1350010 (2013)
\vspace*{.25\baselineskip}\newline
Preprint:
arXiv:1112.2669  
\vspace*{.25\baselineskip}\newline
\vspace*{.25\baselineskip}\newline
\end{abstract}

\maketitle

\section{Introduction}\label{sec:intro}

It is sometimes said that the Standard Model
by itself does not allow for non\-vanishing neutrino masses.
Taking the Standard Model to refer solely to its
particle content and gauge-interaction
structure,\cite{EWSM-a,EWSM-b,EWSM-c,EWSM-d,%
BIM1972-GJ1972-a,BIM1972-GJ1972-b,%
QCD-a,QCD-b,QCD-c,QCD-d} this is not so.
It is, in fact, possible to introduce an interaction term in the
Lagrange density, which uses only the standard-model multiplets
and generates Majorana masses for the neutrinos.
This term is nonrenormalizable and does not conserve $B-L$,
the difference of the baryon quantum number $B$
and the lepton quantum number $L$.

Having a nonrenormalizable theory makes sense as long as the
Standard Model is not considered to be the definite and final theory.
 From this point of view, the interaction term discussed here will
have crossed the mind of anyone who has pondered the origin
of neutrino mass. Indeed, one of the earliest papers to mention this
term appeared more than 30 years ago.\cite{Weinberg1979}
Still, it may be useful to clarify the
basic logic of this term and to emphasize the crucial role
of gauge invariance.

\section{Interaction term}\label{Interaction-term}

The Standard Model of elementary particle physics combines
the chiral $SU(2) \times U(1)$ gauge theory
of the electroweak interactions\cite{EWSM-a,EWSM-b,EWSM-c,EWSM-d}
having anomaly cancellations between the different Weyl fermions
present\cite{BIM1972-GJ1972-a,BIM1972-GJ1972-b}
and
the vectorlike $SU(3)$ gauge theory
of the strong interactions.\cite{QCD-a,QCD-b,QCD-c,QCD-d}
(Further references can be found in, e.g., Ref.~\refcite{ChengLi1984} and
the Feynman rules are given in, for example,
Ref.~\refcite{Veltman1994}.)
The particle content of the minimal Standard Model
consists of the $SU(3) \times SU(2) \times U(1)$
gauge bosons, $N_\text{fam} \times 15=45$ left-handed
Weyl fermions for family number $N_\text{fam}=3$, and a
single complex isodoublet scalar Higgs field. In the following, we
focus on the leptonic sector  (charged leptons $f^\pm$ and    
neutrinos $\nu_f$, with family label $f=e,\,\mu,\,\tau$)
and use the notation of Ref.~\refcite{ChengLi1984}
in terms of four-component Dirac spinors.

The $SU(2)\times U(1)$ irreducible representations
of the lepton families and the Higgs field are
of the type $(\mathbf{2})_Y$ and $(\mathbf{1})_Y$,
that is, isodoublet and isosinglet with $U(1)$ hypercharge $Y$.
Given the definition of the electric charge $Q\equiv I_3+Y/2$,
the basic Weyl (anti-)fermion fields of the first lepton family
(label $f=e$) and the Higgs field are:
\begin{subequations}\label{eq:isodoublets-isosinglets}
\begin{eqnarray}\label{eq:isodoublets-isosinglets-leptons}
L_e &=&
\left(
\begin{array}{c}
\nu_{e,L} \\e^{-}_{L} \\
\end{array}
\right)_{-1}
,\quad
R_{e}  = \big(e^{+}_{R}\big)_{+2} \,,\\[2mm]
\label{eq:isosinglets-Higgs}
\Phi &=&
\left(
\begin{array}{c}
\phi^{+}\\\phi^{0} \\
\end{array}
\right)_{+1},\quad
\widetilde{\Phi} \equiv  i\tau_2\,\cdot\,\Phi^\ast
\equiv
\left(
  \begin{array}{cc}
    0 & \;\;+1 \\
    -1 & \;\;0 \\
  \end{array}
\right)
\cdot\,\Phi^\ast \,,
\end{eqnarray}
\end{subequations}
where the asterisk in the last definition of
\eqref{eq:isosinglets-Higgs} denotes complex conjugation
and the three matrices $\tau_a$
are the standard $2\times 2$ Pauli matrices
for isospin (denoted $\sigma_a$ for spin).
The isodoublet in \eqref{eq:isodoublets-isosinglets-leptons}
has lepton number $L=+1$ and the
isosinglet in \eqref{eq:isodoublets-isosinglets-leptons} 
has $L=-1$.
The (anti-)leptons of the second and third families are
contained in similar representations,
$L_f$ and $R_f$ for label $f=\mu,\,\tau$.
The usual Higgs vacuum-expectation-value constant $v$ is
obtained from $<\Phi^\dagger\cdot\Phi>\,\equiv v^2/2$.
Experimental indications for the existence of a
125 GeV Higgs boson have been reported
recently.\cite{Higgs-ATLAS2012,Higgs-CMS2012}

The generalized theory, now, is defined by the local Lagrange
density $\mathcal{L}_\text{SM}$ 
of the minimal Standard Model,\cite{ChengLi1984,Veltman1994}
to which is added a local lepton-Higgs interaction
term $\mathcal{L}_{5} $,
\begin{subequations}\label{eq:L-L5}
\begin{eqnarray}\label{eq:L}
\mathcal{L}(x) &=&
\mathcal{L}_\text{SM}(x)+\mathcal{L}_{5}(x) \,.
\end{eqnarray}
Specifically, take the following contact-interaction term
which is both $SU(2)\times U(1)$ gauge invariant and
Lorentz invariant  ($\hbar=c=1$):
\begin{eqnarray}\label{eq:L5}
\mathcal{L}_{5}(x)  &=&
\frac{1}{M_{5}}\;\sum_{f,f^{\prime}}\;
\Big[\,\lambda_{f,f^{\prime}}\,
\Big(\,\overline{L}_{f}(x)\cdot \widetilde{\Phi}(x)\,\Big)\;
\Big(\, \widetilde{\Phi}^\dagger(x)\cdot L_{f^{\prime}}(x)\Big)^{c}
+ \text{H.c.}\,\Big] \,,
\end{eqnarray}
\end{subequations}
where  the charge conjugate of the Dirac spinor field $\psi(x)$ is denoted
$\psi^{c}(x)\equiv \mathcal{C}\, \gamma^0\,  \psi^\ast(x)$, with
$(\mathcal{C}\gamma^0) \,  (\gamma^\mu)^\ast\,  (\mathcal{C}\gamma^0)^{-1}=- \gamma^\mu$.
The composite field operator on the right-hand side of
\eqref{eq:L5} has mass dimension five, hence the suffix `5.'
As mentioned before, the dimension-5 term \eqref{eq:L5}  
has already been considered in Ref.~\refcite{Weinberg1979}.

Expanding the Higgs isodoublet $\Phi$ from \eqref{eq:isosinglets-Higgs}   
around its vacuum expectation value $(0,\,v/\sqrt{2}\,)^\text{T}$ gives
\begin{eqnarray}\label{eq:L5-expanded}  
\mathcal{L}_{5}  &=&
\frac{v^2}{2\,M_{5}}\;\Big[\,\sum_{f,f^{\prime}}\; \lambda_{f,f^{\prime}}\;\,
\widehat{\nu}_{f}^\text{\,T}\;(-i\sigma_2)\;
\widehat{\nu}_{f^{\prime}}+ \text{H.c.}\,\Big]+ \,\cdots\, \,,
\end{eqnarray}
where the superscript `$\text{T}$' stands for transposition and
$\widehat{\nu}_{f}$ is the left-handed two-component Weyl spinor
corresponding to the four-component Dirac spinor $\nu_{f,L}$
in the chiral representation of the Dirac gamma matrices,
$\gamma^5\equiv  i\,\gamma^0\gamma^1\gamma^2\gamma^3=\text{diag}(1,\,1,\,-1,\,-1)$.
The first term on the right-hand side of \eqref{eq:L5-expanded}
contains a mix of Majorana mass
terms.\footnote{The manifest $SU(2)$ gauge invariance of \eqref{eq:L5} and
rotation invariance of \eqref{eq:L5-expanded}
rely on identical mathematics: for isospin, the identity
$\Omega^{\dagger}\cdot(i\tau_2)\cdot\Omega^\ast = i\tau_2$
with an arbitrary matrix
$\Omega = \omega_a\, i\tau_a + \omega_4\, \openone  \in SU(2)$
having real parameters $\omega_\mu$ on the unit 4-sphere
[\,$\sum_{a} (\omega_a)^2 +(\omega_4)^2 =1$\,]
and, for spin, the same identity but now in terms of $\sigma_a$.
The transposition and commutation properties of the Pauli matrices
$\sigma_a$ make for the manifest invariance
of \eqref{eq:L5-expanded} under Lorentz boosts.
Remark that also the $U(1)$ gauge invariance of \eqref{eq:L5}
holds separately for the two $L^{\dagger} \cdot \widetilde{\Phi}$-type terms.}  
\textit{A priori}, there is no connection between the neutrino masses
from \eqref{eq:L5-expanded} and the charged-lepton masses.

The ellipsis in \eqref{eq:L5-expanded} contains interaction terms
involving the components of the Higgs isodoublet field. In unitary gauge,
$\Phi(x)=(0,h(x)+v/\sqrt{2}\,)^\text{T}$ with $h(x) \in \mathbb{R}$,
the Feynman rules of the new cubic ($h\,\nu\,\nu$)
and quartic ($h\,h\,\nu\,\nu$)
vertices are obtained from the following Lagrange density:
\begin{eqnarray}\label{eq:L5-unitary-gauge}
\hspace*{-4mm}
\mathcal{L}_{5}^\text{(unitary\;gauge)}  &=&
\frac{1}{M_{5}}
\;\Big( \frac{v^2}{2} + \sqrt{2}\,v\,h + h^2 \;\Big)
\;\Big[\,\sum_{f,f^{\prime}}\; \lambda_{f,f^{\prime}}\;\,
\widehat{\nu}_{f}^\text{\,T}\;(-i\sigma_2)\;
\widehat{\nu}_{f^{\prime}}+ \text{H.c.}\,\Big]\,.
\end{eqnarray}
If nonzero neutrino masses are taken as input
($\lambda_{f,f^{\prime}} \ne 0$),
these new scalar-neutrino interactions are an unavoidable
consequence of our approach and contribute, for example,
to flavor-changing neutrino-neutrino scattering
$\nu_{e}\,\nu_{e} \to \nu_{\mu}\,\nu_{\mu}$
at small but finite center-of-mass energies,
$0 < \sqrt{s} \ll M_5$.
Naive estimates suggest that these new contributions satisfy
the supernova bounds\cite{Manohar1987}
on neutrino-neutrino scattering cross-sections
but definitive statements have to wait for the proper
UV completion of \eqref{eq:L-L5}, as will be discussed in the next section.

\section{Discussion}\label{Discussion}

The interaction term \eqref{eq:L5} is nonrenormalizable
because of the dimensional coupling constant $1/M_{5}$.
This mass scale $M_{5}$ may be related to
the energy scale at which the $B-L$ global symmetry is broken
($B+L$  is already broken dynamically at the electroweak
scale\cite{tHooft1976PRL,KlinkhamerManton1984,KlinkhamerLee2001}).
The experimental data from particle physics and cosmology suggest a
sub-eV neutrino mass scale,\cite{PDG2010} which,
with $v \sim 10^{2}\;\text{GeV}$ and $\lambda_{f,f^{\prime}} \sim 1$
in \eqref{eq:L5-expanded}, implies $M_{5} \gtrsim 10^{13}\;\text{GeV}$.
But $M_{5}$ could also drop to the $\text{TeV}$ scale if, for some reason,
the couplings $\lambda_{f,f^{\prime}}$ were of order $10^{-10}$.

From a purely theoretical point of view, the neutrino mass scale $v^2/M_{5}$
in \eqref{eq:L5-expanded} traces back to the gauge invariance
of \eqref{eq:L5}
[two Higgs isodoublets for the ``saturation'' of the two lepton
isodoublets giving the factor $v^2$]
and nonrenormalizability [giving the factor $1/M_{5}$].
The same type of mass scale $v^2/M_{R}$ follows, of course,
from the see-saw mechanism\cite{see-saw-mech-a,%
see-saw-mech-b,see-saw-mech-c,see-saw-mech-d,see-saw-mech-e}
(brief reviews can be found in 
Refs.~\refcite{ChengLi1984} and ~\refcite{PDG2010}).
The see-saw mechanism, in its simplest form,
introduces $N_\text{fam}$ right-handed neutrinos
[possibly coming from an $SO(10)$ grand unified theory] and
has, per family, an effective $2\times 2$
neutrino-mass matrix with diagonal entries $0$ and $M_{R}$ and
off-diagonal entries $v$ (giving eigenvalues
$M_{R}$ and $-v^2/M_{R}$ for $v^2 \ll M_{R}^2$).
But, here, there are no right-handed neutrino fields
and there is no such $2N_\text{fam}\times 2N_\text{fam}$
matrix to diagonalize, only the $N_\text{fam}\times N_\text{fam}$
matrix from \eqref{eq:L5-expanded} with entries
individually of order $v^2/M_{5}$.

In the context of renormalizable theories,
there are also alternatives to heavy right-handed neutrinos;
see, in particular, the discussion of Ref.~\refcite{Ma1998}.
These different realizations can be expected to lead
to different results for the neutrino-neutrino scattering cross-sections
discussed in the last paragraph of Sec.~\ref{Interaction-term}.
(For neutrino-neutrino scattering,
certain statements in  Ref.~\refcite{Ma1998}
as to the indistinguishability of the different
realizations presumably hold only in the
strict low-energy limit, $\sqrt{s}/M_5 \to 0$.)

Let us make two final comments.
First, it is remarkable that all experimental facts of elementary
particle physics known to date\cite{PDG2010}
can be described precisely by the fermion and Higgs multiplets
of the minimal Standard Model if one allows for a single
nonrenormalizable contact-interaction term in the action.
These experimental facts include those from
neutrino oscillations and perhaps those from neutrino-less
double-beta decay. In principle, even Lorentz-violating effects
could be
incorporated.\footnote{A hypothetical superluminal neutrino velocity
(first claimed to have been discovered by OPERA
in Ref.~\refcite{OPERA2011-a},
but later retracted in Ref.~\refcite{OPERA2011-b}) could  also be
described by using only the  multiplets of the minimal Standard Model.
The idea would be to appeal to
spontaneous breaking of Lorentz
invariance\cite{SBLI-a,SBLI-b,SBLI-b}
and to take a Majorana-mass-type interaction term as \eqref{eq:L5}
with derivatives inserted between 
the two gauge-invariant $L^{\dagger} \cdot \widetilde{\Phi}$-type terms  
(these derivatives are to be contracted with condensate vectors or tensors).}  

Second, the origin of the term \eqref{eq:L5} may very well rely on an
explanation which does not involve right-handed neutrinos or even
grand unification. In fact, it could be that
the apparent merging of the running $SU(3) \times SU(2) \times U(1)$
gauge coupling constants at high energies ($E \sim 10^{15}\;\text{GeV}$)
would not signal the appearance of a unified gauge 
group\footnote{Without
the simple gauge group of grand unification, for example $SO(10)$,
there would be 
no proton decay and no magnetic-monopole soliton solution
(for original references and brief reviews on both topics,
see, e.g., Ref.~\refcite{ChengLi1984}).} 
but the onset of nonperturbative dynamics responsible for 
compositeness of the gauge bosons\cite{KlinkhamerVolovik2005} 
[an alternative scenario relies on a Lorentz-violating        
deformation of chiral gauge theory\cite{Kawamura2009}].        
A further consequence of these new underlying interactions might then be
the appearance of an effective interaction term \eqref{eq:L5} which violates
$B-L$, in addition to the $B+L$ violation
inherent to the electroweak chiral gauge
theory.\cite{tHooft1976PRL,KlinkhamerManton1984,KlinkhamerLee2001}
In this way, the conservation of both
baryon number $B=(B+L)/2+(B-L)/2$ and lepton number $L=(B+L)/2-(B-L)/2$
would be only approximate at low
energies, because the fundamental fermionic constituents would not
care about these quantum numbers.
The $SU(3) \times SU(2) \times U(1)$ gauge symmetry would
be an emergent symmetry
and the small neutrino mass scale would be a remnant of 
such a state of affairs.


\end{document}